\documentstyle[aps,prl,epsfig]{revtex}

\title{Dipoles and fractional quantum Hall masses}
\begin{document}
\author{V. Pasquier}
\address{Service de $%
\ddot{}%
$Physique Th\'{e}orique, CEA/Saclay, F-91191 Gif-sur-Yvette Cedex, FRANCE}
\date{\today}
\maketitle
\tightenlines
\widetext
\begin{abstract}
We develop a microscopic formalism to study the fractional quantum Hall
plateaus at filling factors $\nu $ away from $1/2\beta$  $\beta$ an integer.
The theory is in terms
of quasiparticles which carry a charge $e^{\ast }$ equal to $1-2\beta\nu $ times
the charge of the electron. The wave functions obtained following our
approach are shown to coincide precisely with the form predicted by Jain and
this holds independently of the interaction potential.\ Microscopically this
rigidity originates from the fact that two different charges interacting
attractively in their lowest Landau levels form a bound state with a
universal wave function.\ From the expressions of the gaps we compute an
effective mass which agrees well with the experiments carried at
$\nu=1/2$ and $1/4$.

\end{abstract}


Many fundamental aspects of
the fractional quantum Hall effect (FQHE) have resulted from a better understanding
of the vicinity of the $\nu=1/2$ filling factor. 
Jain was able to obtain the most prominent FQHE plateaus
by reinterpreting the FQHE as an integer quantum Hall effect (IQHE) 
for particles which experience a reduced
magnetic field \cite{JAIN}. 
Halperin Lee and Read have predicted that the system  behaves in a Fermi liquid way \cite{HAL} 
at the filling fraction $\nu=1/2\beta,\ \beta=1,2$. Their theory
provides a convincing explanation of the anomaly observed at $\nu=1/2$ 
(and less strongly
at $\nu=1/4 $) 
in the surface acoustic
propagation
by Willet et al. \cite{WIL}.
Several experiments \cite{EXP} have seen that slightly away from $\nu =1/2\beta$ 
the electrons move in circles as if they experience a reduced
magnetic field $\Delta B=B-B_0$.
A phenomenological way to account for these properties is to
assume that the relevant excitations, also 
called composite fermions (CF), carry a fractional charge varying
continuously with the filling factor, $e^{\ast }=(1-2\beta\nu )e$. 
Although such fractional charges have been observed in
noise experiments \cite{GLA}, they have not been clearly identified
with the CF. We
propose here a microscopic derivation of the CF properties in
terms of quasiparticles which carry a fractional charge.
Each electron is replaced by a quasiparticle and the missing charge is
carried by an incompressible charged vacuum.

Another feature of the FQHE is that  
unlike in the IQHE, the gaps responsible for it
are  due to
the interactions. 
Here we set mass of the electrons to zero
and the CF acquire an effective cyclotron
energy due to their interactions. The novelty 
is to predict microscopically a rigid structure for the wave functions 
independently of the interacting potential. 
When the filling fraction gets
close to $1/2\beta$ the states connect continuously to
the Fermi-liquid states at $\nu =1/2\beta$.
The activation gaps are then given by the Landau Level spacing
$\Delta=e^*\hbar B/m^*$ where $m^*$ defines the composite-fermions effective mass.
The masses we obtain are very sensitive to the parameter $\lambda$
which simulates
the effect of the thickness of the sample.
A very good agreement with experimental results \cite{MASS}
is obtained when this parameter is of the order of the magnetic length 
\cite{MORF}.

In order to understand the origin of the effective charge $e^{\ast }$ and how the
cyclotron energy results from the repulsive interactions one approach
initiated by Read \cite{READ} (see also \cite{PAS,SHA,LEE,VON}) suggests
that the $\nu =1/2$ low energy quasiparticles are dipoles. This picture can
be related to the trial wave functions \cite{JAIN,REZ,HAL} if 
one notes \cite{HAL,PAS,READ2,KAM}
that the effect of the Lowest Landau Level (LLL) projection 
inherent to these wave functions is to displace the electrons
from their positions in a $\nu =1/2\beta$ bosonic Laughlin wave
function. In \cite{PAS,PAHA} we developed a theory where the fundamental
excitations were these dipoles.
We represented one excitation
by two opposite charges interacting attractively in
a magnetic field \cite{GOR,LER,KAL}. They form a bound state which moves
in a straight line and carries a dipole momentum perpendicular to its
canonical momentum. When projected into the LLL the eigenfunctions are
independent of the attractive potential. They can be expanded on the LLL
basis of the particle and hole and the expansion coefficients form a matrix
which is the projection of a plane wave into the LLL. To a first
approximation the $\nu =1/2\beta$ CF can be described by non interacting dipoles
and the wave function studied in \cite{REZ,HAL} is the antisymmetrized
product of these projected plane waves.

We follow the same approach to study the CF at filling fractions away from $%
\nu=1/2\beta$. 
The main modification is that the electron charge differs from that of the hole.
The charge of the hole  
varies with the magnetic field  so that 
a system of particles with this charge and a
density equal to that of the electrons
has a constant filling factor equal to $1/2\beta$. 
The reason for this choice is to construct a bosonic system 
which remains incompressible
as the magnetic field is varied and which defines the vacuum of the CF excitations.
In this description the CF is a particle hole excitation where the 
statistics and the charge of the particle differ from that of the hole.
As a consequence
the bound state made by the electron and its correlation hole acquires a charge 
$e^*=(1-2\nu )e$ \cite{PAS,SHA2,FEL}. 
In the following we use units where $\hbar=c=l=1$, $l=\sqrt{\hbar c/eB}$
is the magnetic length. The product of the charge of the electron and of the hole 
by the magnetic field $B$ are denoted $q_a$
and $q_b$.
$q_a=1$ 
in our units
and  $q_b=2\beta \nu q_a$. 

Let us first consider a simple model which contains in essence the
main points developed in this letter:
A single electron and its correlated hole are
coupled by a spring. Their coordinates 
are $x_{a(b)},y_{a(b)}=\vec x_{a(b)}$. In the Landau gauge 
their dynamics follows the Lagrangian:
\begin{eqnarray}
{\it L}=(q_a x_a \dot y_a - q_b x_b \dot y_b)
-K/2( (x_a-x_b)^2+(y_a-y_b)^2)
\label{LAG}
\end{eqnarray}
where in (\ref{LAG}) we have have taken the strong $B$ field limit which enables to
neglect the masses of the particles to be
neglected.
For simplicity we consider here the case $q_a>q_b$.
The  momenta of $y_a$ and $y_b$ are $p_a=q_ax_a$ and $p_b=-q_bx_b$.
The Hamiltonian is: 
\begin{eqnarray}
H=K/2( (p_a/q_a+p_b/q_b)^2+(y_a-y_b)^2)
\label{HAL}
\end{eqnarray}
Its eigenfunctions are given by:
\begin{eqnarray}
\chi_{P_Y,n}(y_a,y_b)=
e^{iP_Y(q_ay_a-q_by_b)/(q_a-q_b)}e^{-(y_a-y_b)^2/(q_b^{-1}-q_a^{-1})}
H_n((y_a-y_b)/(q_b^{-1}-q_a^{-1})^{1/2})
\label{LAH}
\end{eqnarray}
where $P_Y$ is the momentum in the $y$ direction
and $H_n(x)$ are the Hermite polynomials.
The $x$ coordinate 
$X=(q_a  x_{a}+q_b  x_{b})/(q_a+q_b)$ and the $y$-momentum
$ P_Y= p_a + p_b$  are related by $P_Y=\langle X \rangle (q_a-q_b)$.
This means that the two charges
move as a charge $q^*=q_a-q_b$ in the magnetic field. 
The harmonic oscillator frequency  
gives the effective mass of the bound state  to be:
$q_aq_b/K$. Assuming that the electrons behave as a gas
of noninteracting such bound states, at the filling factors of 
the principal series $\nu=n/{2\beta n+1}$
the effective filling factor of the reduced charge $q^*$ is equal to $n$,
which explains the occurrence of the FQHE at these fractions.
At $\nu=1/2\beta$, $q^*=0$, and
the Hamiltonian reduces to a free Hamiltonian $H=P^2/{2m^*}$ so that
we expect a Fermi sea to form.
The momentum $\vec P$ and the relative coordinate are then related by 
$-P_X=q_a(y_a-y_b),\ P_Y=q_a(x_a-x_b)$ so that the bound state
behaves as a neutral dipole with a dipole vector perpendicular and proportional
to its momentum.
Note 
that since the strength $K$ of the spring enters the Hamiltonian (\ref{HAL}) as a 
normalization factor
the wave function (\ref{LAH}) which describes the two charges is independent of $K$.
 
If we replace the spring by a rotation
invariant potential $V(r)$, one can refine the preceding approach to show that
the wave function of the bound state is independent of
the potential. To see it we need to consider the 
problem of the particle and the hole interacting in their
respective LLL.
The 
orbital of a charge $q$ particle in the $n^{th} \ge 0$
Landau level
is denoted $u_{n,t}^{q}$ (simply $%
u_{t}^{q}$ for $n=0$) where  $t$ is the
momentum in the $y$ direction:
\begin{eqnarray}
u_{n,t}^{q}(x,y)=e^{ity}H_{n}(\sqrt{q}(x-t/q))
e^{-q(x-t/q)^2/2}
\label{GEN}
\end{eqnarray}
The  wave function for the  charge and the hole can 
be decomposed on a basis of LLL orbitals:
\begin{eqnarray}
\chi_{P_Y,n}(x_a,x_b,y_a,y_b)=\sum_{s-t=P_Y}  \hat \chi^{P_Y,n}_{st}  
u^{q_{a}}_s(x_a,y_a) \bar u^{q_{b}}_t(x_b,y_b)
\label{LAI}
\end{eqnarray}
and the coefficients $\hat \chi^{P_Y,n}_{st}$ are obtained
by diagonalizing the potential interaction:
\begin{eqnarray}
\sum_{s^{\prime}t^{\prime}} \langle st^{\prime}|V|s^{\prime}t\rangle 
\hat \chi
^{P_Y,n}_{s^{\prime}t^{\prime}}=- \epsilon_{n}^0 \hat \chi^{P_Y,n}_{st}
\label{SECU}
\end{eqnarray}
The matrix elements $\langle st|V|s^{\prime}t^{\prime}\rangle$
are taken between LLL orbitals of charge $q_a$ for $s,s^{\prime}$ and $q_b$
for $t,t^{\prime}$. 
Without loss of generality we restrict ourselves to the
zero momentum case $P_Y=0$, $(s=t,s'=t'),$
and we consider a Gaussian potential $V(r)=e^{-r^2/{2\mu^2}}$.
In the large system size limit, the momentum index $s$ becomes continuous
and the secular problem (\ref{SECU}) rewrites:
\begin{eqnarray}
&&{1\over 1+1/{\mu^2 q_b}}\sqrt{1/q_b-1/q_a\over \pi(1-\tau^2)}
\int_{-\infty}^{\infty}
 e^{-(1/q_b-1/q_a)((s^2+s'^2)(1+\tau^2)/2-2\tau ss')/(1-\tau^2)}
 \chi^n(s') ds'=- \epsilon_{n}^0 \hat \chi^{n}(s) \nonumber \\
&& {\rm with}\  \tau=(1+1/{\mu^2 q_a})/(1+1/{\mu^2 q_b})
\label{SECO}
\end{eqnarray}  

One recognizes the Poisson kernel for the Hermite polynomials.
The eigenfunctions $\chi^{n}(s)$ and eigenvalues $\epsilon_{n}^0$
are thus given by:
\begin{eqnarray}
&&\chi^n(s)=
H_{n}((1/q_b-1/q_a)^{1/2}s)
e^{-(1/q_b-1/q_a)s^2/2}\nonumber \\
&&\epsilon_{n}^0=\tau^n/{(1+1/{\mu^2 q_b})}
\end{eqnarray}
The general eigenvector for $P_Y\ne 0$ 
can be put in the gauge independent form:
\begin{eqnarray}
&\hat \chi^{n,P_Y}_{st}= \langle s|u_{n,P_Y}^{q^*}|t\rangle  
\label{SOLA}
\end{eqnarray}
Here as earlier the momentum $s$ and $t$ refer
respectively to charge $q_a$ and $q_b$ orbitals in the LLL.
This is our main result: 
the wave function of the bound state is universal and given by the projection
of higher Landau levels in the LLL. 
The eigenvalues give the kinetic energy of the bound state and since
the eigenvectors do not depend on $\mu$, they can be obtained
for an arbitrary potential using a superposition principle.
In the case of the
potential \cite{ZDS,YO} $V_{\lambda}(r)=1/{\sqrt{ r^2+\lambda^2}}$
one has:
\begin{eqnarray}
V_{\lambda}(r)&=&\sqrt{2\over \pi}\int_{0}^{+\infty} 
e^{-(r^2+\lambda^2)/{2\mu^2}} {d\mu \over {\mu^2}}\nonumber \\
\epsilon_{n}^0&=&-
\sqrt{2\over \pi}\int_{0}^{+\infty}
e^{-\lambda^2/{2\mu^2}}
{(1+1/{\mu^2 q_a})^n\over (1+1/{\mu^2 q_b})^{n+1}}
{d\mu\over {\mu^2}}
\label{ENO}
\end{eqnarray} 
If we consider the filling factors $\nu_n=n/{2\beta n+1}$
one has $q_a=1,\ q_b=2\beta n/{2\beta n+1}$.
In the absence of interactions the gap $\Delta_{\nu_n}$
is the energy needed to put a particle from the last occupied LL into
the first empty one $\Delta_{\nu_n}=\epsilon^0_n-\epsilon^0_{n-1}$.
In the limit where $n$ is large the gaps become  proportional to the charge 
$q_n^*=q_a-q_b$
and one can reinterpret them as an effective cyclotron energy 
for a particle with an
effective mass $\alpha^0$, $\Delta_{\nu_n}=q^*/\alpha^0$. 
One obtains the following
expression for the effective mass:
\begin{eqnarray}
{1\over \alpha^{0}}=\sqrt{2\over\pi}\int_0^{\infty} e^{-{1\over2\beta(1+\mu^2)}
-{\lambda^2\over 2 \mu^2}}
{d\mu \over (1+\mu^2)^2}
\label{MAO}
\end{eqnarray} 
We can
see the main features of the inverse-mass behavior in this expression.
It decreases mildly with the filling factor $1/2\beta$ and 
drastically with
the thickness parameter $\lambda$. In the following we improve
this formula to take into account the interactions between the CF.

The characteristic feature of the orbitals 
(\ref{SOLA}) is that they contain no adjustable parameters and
we expect that the wave 
functions obtained by antisymmetrizing them
coincide with the trial wave functions \cite{KAM,REZ} which are by construction
independent of the interactions.
Nevertheless, in order to obtain quantitative results
we need to determine the coupling between the hole and the electron
in a consistent way. 
To reach this goal 
we use a second quantized formalism.
The bosonic ($b_{t}^{+},b_{t}$) and fermionic ($a_{s}^{+},a_{s}$%
) creation and annihilation operators create respectively a charge $%
q_{b}$ boson with the y-momentum equal to $t$ and a charge $q_{a}$ electron
with the y-momentum equal to $s$ in the LLL. The fields $\Phi _{b}(\underline{x}%
)=\sum_{t}u_{t}^{q_{b}}(\underline{x})b_{t}$ ($\Phi _{a}(\underline{x}%
)=\sum_{s}u_{s}^{q_{a}}(\underline{x})a_{s}$) and its hermitian conjugate
annihilate and create a boson (an electron) at position $\underline{x}$.
The field  $\psi ^{+}(\underline{x})=\Phi _{a}^+\Phi _{b}(\underline{x})$  
removes a boson of charge $q_b$  at position $\underline{x}$
and replaces it with a fermion with charge $q_a$. 
The state $|\Omega \rangle$ defines the ground state
of $N$ bosons 
at the filling factor $2\beta^{-1}$.
If we act on $|\Omega \rangle $ with $N$ creation operators $\psi ^{+}(\underline{x})$
the resulting state contains only electrons.
The idea which motivates the following construction is that the state 
$|\Omega \rangle$ is rigid against a density wave excitation
so that the low energy physics   
results from the dynamics of the field $\psi ^{+}(\underline{x})$.
When we act on $|\Omega \rangle$ with $\psi ^{+}(\underline{x})$ we assume that the separation
of the electron and its correlation hole remains sufficiently small so that we can treat the
pair as a bound state.
We derive an effective
Hamiltonian for these excitations which we break into a kinetic term and an interaction potential
and we recover the wave functions of the bound states as the eigenstates of the
kinetic term. 

Let $\rho(\underline{x})$ denote the electron density. The dynamics of the electrons is
governed by the Hamiltonian:
\begin{eqnarray}
H=1/{2}\int d^2\underline x d^2\underline y V_{e}(\underline x-\underline y)
\rho(\underline x)\rho(\underline y) \label{HAMI}
\end{eqnarray}
where $V_{e}(\underline x)$ is the electron-electron potential.
In the operatorial approach 
the coordinate of CF operators acting on $|\Omega \rangle $ coincide with 
the position of the electrons and
we can identify $\rho$ with  the CF density operator $\rho_a=\Phi _{a}^+\Phi _{a}$.
Since there is no boson in the physical system 
the bosonic density $\rho_b=\Phi _{b}^+\Phi _{b}$ is equal to zero.
Therefore, if we substitute $\rho$ with $\rho_a +\kappa \rho_b$ in (\ref{HAMI})
the quantities we compute with this modified density are in principle independent of $\kappa$.
The scheme we are using, however,  does not allow us to set $\kappa$ to $0$
because the potential which binds the particle to
the hole is equal to $-\kappa V_{e}(\underline x-\underline y)$ and 
our approach thus requires a positive value of $\kappa$ to bind them. 
We can determine the best value for $\kappa$ 
so that the derivative of a physical quantity
with respect to $\kappa$ is zero.
We shall therefore work out the formalism 
using the above expression of $\rho$ in (\ref{HAMI}) and 
we shall set the
value of $\kappa$ 
using this procedure when we compute
the gaps.
Alternatively, Shankar \cite{SHA2,MUR} has suggested to determine $\kappa$
by letting the bosonic and
fermionic particles interact proportionally to their charge
and set $\rho(\underline x)=
q_a\rho^a(\underline x)+q_b\rho^b(\underline x)$ (in the $\nu=1/2\beta$
case this coincides with our proposal \cite{PAS,PAHA}). 
Although both procedures give comparable results at $\nu=1/2\beta$,
the first one is more reliable away from this filling factor.

To recast the dynamics in terms of CF  
it is useful to replace the composite operator $a^+_sb_{t}$
by a fermionic creation operator $\psi^+_{st}$.
For this, we use the fact that the
densities $\rho^a_{ss^{\prime}}=a^+_sa_{s'}$ and $\rho^b_{tt^{\prime}}=b^+_tb_{t'}$ 
entering the definition of $\rho$ in (\ref
{HAMI}) are the generators 
of two commuting algebras $U(q_a)$ and $U(q_b)$ which have a natural
representation in terms of matrix operators $\psi^+_{st}$. We
define the CF creation and annihilation operators as a set of matrix
fermions: 
\begin{eqnarray}
&&\{\psi_{ts},\psi_{t^{\prime}s^{\prime}}\}=
\{\psi^+_{st},\psi^+_{s^{\prime}t^{\prime}}\}=0  \nonumber \\
&&\{\psi^+_{st},\psi_{t^{\prime}s^{\prime}}\}=\delta_{tt^{\prime}}
\delta_{ss^{\prime}}  \label{FERMCO}
\end{eqnarray}
and we represent the generators as: 
\begin{eqnarray}
\pmatrix{a^+_sa_{s'}&a^+_sb_{t'}\cr b^+_ta_{s'}&b^+_tb_{t'}\cr} = %
\pmatrix{(\psi^+\psi)_{ss'}& \psi^+_{st'}\cr
(\psi\psi^+\psi)_{ts'}&(\psi\psi^+)_{tt'}\cr}  \label{HOLS}
\end{eqnarray}
It is easy to verify that the commutation relations of the matrix elements of
the right and left matrix coincide and in fact, the right hand side generators
can be viewed as a generalized Holstein-Primakov transformation for the
left hand side generators.
More familiar generators are given by the Fourier modes 
of the densities $\rho^a(\underline x)$ and $\rho^b(\underline x)$
which generate the two Girvin, MacDonald and Platzman
algebras of the Fermionic and Bosonic densities \cite{GIR}.

The effective Hamiltonian  
is obtained by substituting the densities $\rho(%
\underline x)$ with their expression in terms of CF operators
(\ref{FERMCO})
in the Hamiltonian (\ref{HAMI}). 
In order to be useful this description requires 
the use of two approximations.
The first one assumes the vacuum 
$|\Omega\rangle$ is annihilated by the CF operators $\chi_{ts^{\prime}}$.
It is justified if the gap a bosonic system at
the filling factor $1/2\beta$ is considerably larger than the physical gap
of the electron system 
and we can disregard
the fluctuations of the density ($\rho_b$) in this state.
In \cite{PAS,PAHA} we considered a
system of bosons at $\nu=1$ $(2\beta=1)$ where this condition is exactly realized.
The second
approximation consists in identifying the physical Hilbert space generated by $%
a^+_sb_t$ with the over-complete space generated by $\psi^+_{st}$.
(It has been related to a gauge invariance by Read in \cite{READ2}). 
A consequence is that in the CF description physical quantities depend on the bosonic
interactions through the parameter $\kappa$. 
As a consistency check we shall verify that
the value of $\kappa$ which minimizes the gaps is such that the
kinetic energy contribution is large compared to the interactions.
 
To obtain the physical CF creation operators and their kinetic energy 
we diagonalize the Hamiltonian in the one particle Hilbert space. 
If we denote by
$\psi_{n,P_y}^+= \sum_{st} \hat \chi^{n,P_y}_{st} \psi^+_{s,t}|\Omega\rangle$
the eigenstates, the secular problem coincides precisely with (\ref{SECU}).
The eigenstates of the kinetic energy are thus generated upon acting
on the vacuum $|\Omega\rangle$ with $N$ operators $\psi^+_{n,P_y}$
with $\hat \chi^{n,P_y}_{st}$ given by (\ref{SOLA}).
It is now straightforward to verify that the ground state wave function 
takes the form:
\begin{eqnarray}
\Psi (\underline x_i) ={\cal P}^{(q_a)} \Phi_n^{(q^*)}(\underline x_i)\
\Phi^{(q_b)}_{\Omega}(\underline x_i)  \label{JAIN}
\end{eqnarray}
where $\Phi^{(q_b)}_{\Omega}(\underline x_i)$ is the ground state wave function
for 
$N$ bosons of charge $q_b$ at the filling factor $%
1/2\beta$ and $\Phi_n^{(q^*)}(\underline x_i)$ is the Slater determinant of
$N$ charge $q^*$ orbitals filling $p$ Landau levels.
The symbol ${\cal P}^{(q_a)}$ means that we expand the total wave function on
the basis of Slater determinants of $N$ charge $q_a$
orbitals belonging to all possible LL
and we project it on the subspace of determinants with all their
orbitals in the LLL. It has the effect
of replacing the higher orbitals $u^{q^*}_{n,t}$ appearing in the factor $\Phi_n^{(q^*)}$ by
their matrix elements (\ref{SOLA}) between LLL orbitals of charge $q_a$ and $q_b$.
This projection differs slightly
from the one in \cite{KAM} which is defined to act separately
on different factors of the total wave function.   
           
The CF excited states can be described 
in a similar way as the states of a Fermi liquid theory.
We  put the system in a box of area $2\pi/q^*$ by keeping only the CF orbitals $u^{q^*}_{n,t}$ 
with the position index $t$ equal to $0$.
In presence of a magnetic field, the momenta of the CF must be replaced by
the pseudo-momenta:
$\pi_x= p_x,\ \pi_y= p_y+q^* x$
which not commute and therefore cannot be diagonalized simultaneously.
We can still define
coherent states $|k\rangle$ which 
diagonalize the complex pseudomomenta $\pi_-=\pi_x -i\pi_y$:
\begin{eqnarray}
\langle \underline x| k\rangle =e^{-{\frac{\bar k k }{4 q^*}}} \sum_{m=0}^\infty {%
\frac{1}{m!}} ({\frac{k}{\sqrt{2q^*}}})^m  u^{q^*}_{m,0}(\underline x)
\label{COH}
\end{eqnarray}
An excited state is characterized by the occupation number $n(k)$
of the different coherent states $|k\rangle$. The energy per
unit area is obtained by taking the expectation value of the Hamiltonian (\ref{HAMI}) in this state : 
\begin{eqnarray}
E_{\{n\}}={\frac{2\pi}{q^*}} \sum_{k,k^{\prime},p} n(k)(1-n(%
k^{\prime})) \hat V(\underline p) \langle\bar k|\rho(\underline p%
)|k^{\prime}\rangle\langle\bar k^{\prime}|\rho(-\underline p)|k\rangle\cr
\label{HFE}
\end{eqnarray}
where the density $\rho=\rho^a+\kappa\rho^b$ and the matrix elements of  
$\rho^a$ and $\rho^b$
in the coherent states are given by: 
\begin{eqnarray}
\langle\bar k|\rho^b(\underline p)|k^{\prime}\rangle= e^{-(p\bar p(1+2{\frac{%
q^*}{q_b}})+k\bar k +k^{\prime}\bar k^{\prime}-2k\bar k^{\prime}-2{\frac{%
\sqrt{q_a}}{\sqrt{q_b}}}(\bar k p-k^{\prime}\bar p))/{4q^*}}\cr \langle\bar k%
|\rho^a(\underline p)|k^{\prime}\rangle= e^{-(p\bar p+k\bar k +k^{\prime}%
\bar k^{\prime}-2k\bar k^{\prime}-2{\frac{\sqrt{q_b}}{\sqrt{q_a}}}(\bar k
p-k^{\prime}\bar p))/{4q^*}}\cr  
\label{RHO}
\end{eqnarray}
The energy of a quasiparticle with pseudo-momentum $k$ ,$\epsilon(k)$, is obtained
by differentiating $E_{\{n\}}$ with respect to $n(k)$.
When the filling factor $\nu_n=n/(2n\beta+1)$ gets closed to $1/2\beta$,
the effective charge $q^*\to 0$ and the pseudo-momentum $%
k$ gets peaked around the level $m$ where $m/n=|{\frac{k}{k_f}}|^2$
and $k_f=\sqrt{1/\beta}$. In this limit, the ground state coincides with the state
for which $n(k)=1$ for $|k|<k_f$ so that
$k_f$ defines the Fermi momentum.
We can also verify that the matrix elements (\ref{RHO}) yield back the conservation of the momentum
$\delta(\underline k-\underline p-\underline k')$
and the expression of the
energy (\ref{HFE}) converges towards the HF energy of the Fermi liquid at $%
\nu=1/{2\beta}$ \cite{PAS}: 
\begin{eqnarray}
E_{\{n\}}&=&{\frac{1}{2}}\sum_{k,k^{\prime}} n(\underline k) (1-n(\underline %
k^{\prime})) \tilde V(\underline k-\underline k^{\prime})
(1+\kappa^2-2\kappa\cos (\underline k \times \underline k^{\prime})) \nonumber \\
{\rm with}&& \tilde V(\underline k) =e^{- k^2/2}\hat V(\underline k)
\label{ENER}
\end{eqnarray}
The effective mass of the Fermi liquid is given by: $k_f/\alpha={%
\frac{\partial \epsilon }{\partial k}}(k_f)$.
This gives a quadratic expression in the variational parameter $\kappa$
which we minimize with respect to $\kappa$.

The inverse masses $1/\alpha$   are plotted in Fig.1 for the potential 
$V_{\lambda}(r)=1/\sqrt{\lambda^2+r^2}$ 
and for the
two filling factors   $\nu=1/2,1/4$.

\begin{figure}[ht]
\begin{center}
\epsfig{file=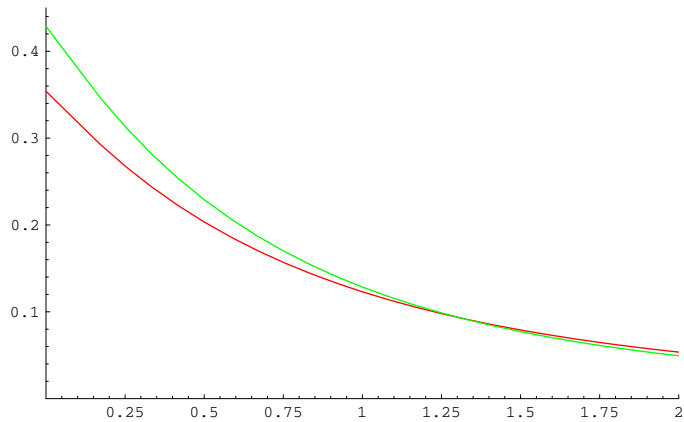,width=10cm}
\end{center}
\caption{The inverse mass $1/\alpha$ is plotted versus $\lambda$ for the two filling
factors $\nu=1/2$ (bottom) and $\nu=1/4$ (top)}
\end{figure}

As  a consistency test of the approximation we can verify that
the parameter $\kappa$ adjusts its value so that the effect of the interactions is small.
Indeed the ratio $\alpha/\alpha^{0}$ where $\alpha^{0}$
is given by (\ref{MAO}) has its minimum  $.8$ for $\lambda=0$
and increases very fast to $1$.
The Coulomb inverse mass ($\lambda=0$) at $\nu=1/2$ is slightly larger
than the prediction of Jain and Kamilla \cite{KAM}: $1/\alpha=.31$.
The gaps and the inverse mass
decrease very fast with the thickness parameter $\lambda$
which reflects the fact that they are mainly due to the part of the
potential of the order of the magnetic length.
In order to compare this results with experiments we express the
ratio of the effective mass to the bare mass of the electron as \cite{PARK}:
\begin{eqnarray}
{m^*\over m_e}={\hbar \omega \over V_c}
\end{eqnarray}
$\hbar \omega=\hbar e B/m_e c$ 
is the cyclotron energy of a bare electron 
and $V_c=e^2/\epsilon l$ is the energy scale of the problem ($\epsilon=12.8$ is the dielectric
constant for GaAs). It gives ${m^*\over m_e}=2.63\ 10^{-2 }\alpha\sqrt{B}$ ($B$ in tesla). 
A value of $m^*=0.26 \sqrt{B}$  in good agreement with the 
experimental observations \cite{MASS}
is obtained for $\lambda=1.2$ (in units of $l$). This value has been
argued by Morf \cite{MORF} to  be a reasonable choice.
For this value of $\lambda$
the masses at $\nu=1/4$  and $\nu=1/2$ differ very little, in agreement
with Pan et al. \cite{WEST}.

In conclusion we have proposed a simple analytical model to study the
FQHE gaps in the vicinity of fillings $\nu=1/2\beta$ with $\beta$ an integer.
Its main virtue is to predict microscopically wave functions
with no
adjustable parameter. 
Two main feature are predicted.
The composite fermion mass depends little on the filling fraction $1/{2\beta}$ as seen in \cite{WEST} and
the gaps collapse with the thickness parameter $\lambda$. 
The first feature is easy to understand: The filling fraction enters only the definition
the Fermi momentum $k_f=1/\sqrt{\beta}$ and as long as the CF intraction is weak, the mass depends
little on $k_f$. The second feature raises questions when we compare this theory to the
scaling $m^*\propto \sqrt{B}$ seen in \cite{WEST}. The Thickness parameter $\Lambda=\lambda l$
introduces a new length scale and we see only two ways to explain this scaling.
Either $\lambda$ is independent of the density (in other words $\lambda$ scales like $l$)
which needs to be explained or the mass depends little on $\lambda$ which contradicts our predictions.
We note however that a strong dependence of the gaps in  $\Lambda$ 
may be related to the experimental observation of Shayegan et al \cite{SHAY}.



\vfill\eject

\end{document}